\documentclass{iau_sj}
\usepackage{graphicx}

\newcommand{\oiii}{[\textrm{O}~\textsc{iii}]}

\title[AGN Absorption] 
{AGN Absorption Linked to Host Galaxies}

\author[Juneau et al.]   
{St\'ephanie Juneau
}

\affiliation{CEA-Saclay \\ DSM/IRFU/SAp \\
91191 Gif-sur-Yvette, France \\ email: {\tt stephanie.juneau@cea.fr} \\[\affilskip]
}

\pubyear{2013}
\volume{304}  
\pagerange{--}
\jname{Multiwavelength AGN Surveys and Studies}
\editors{A. Mickaelian, D. Sanders \& F. Aharonian, eds.}
\begin{document}

\maketitle

\begin{abstract}
Multiwavelength identification of AGN is crucial not only to obtain a more complete census, but also to learn about the physical state of the nuclear activity (obscuration, efficiency, etc.).  A panchromatic strategy plays an especially important role when the host galaxies are star-forming. Selecting far-Infrared galaxies at $0.3<z<1$, and using AGN tracers in the X-ray, optical spectra, mid-infrared, and radio regimes, we found a twice higher AGN fraction than previous studies, thanks to the combined AGN identification methods and in particular the recent Mass-Excitation (MEx) diagnostic diagram.  We furthermore find an intriguing relation between AGN X-ray absorption and the specific star formation rate (sSFR) of the host galaxies, indicating a physical link between X-ray absorption and either the gas fraction or the gas geometry in the hosts.  These findings have implications for our current understanding of both the AGN unification model and the nature of the black hole-galaxy connection.
\end{abstract}

\firstsection 
\section{Introduction}

The tight relation between the mass of supermassive black holes (SMBHs) and their host galaxy bulges, and the similarities in the cosmic history of star formation and SMBH growth, motivated several attempts to elucidate the underlying physical connections.  Moreover, AGN feedback has been invoked to regulate star formation in cosmological models of galaxy evolution \cite[(Croton \etal\ 2006)]{cro06}.  However, the effects of AGN on star formation, or conversely, the effect of star formation onto AGN triggering have remained uncertain due to the difficulties in gathering complete samples of AGN.  While there are many AGN selection methods, each single method suffers from incompleteness.  For example, hard X-rays (2-10\,keV) are a reliable signature of AGN but can be absorbed by large column densities of intervening gas ($N_H>10^{23-24}$\,cm$^{-2}$). 

According to AGN unification models, obscuration of the central X-ray emission and of the accretion disk takes place on small scales by a {\it torus} of obscuring material. The orientation of the torus opening with respect to the observer's line-of-sight will determine whether the central regions are directly observable or obscured.  There are several successes of this type of models, but also some shortcomings (e.g., \cite{ale12}). Alternatively, there are expectations for obscuration to take place on larger scales during particularly gas-rich phases of galaxy evolution such as gas-rich major mergers \cite[(Sanders 1988)]{san88}.  According to these scenarios, the central AGN is heavily obscured by surrounding gas and dust during a deeply buried phase, before being visible in X-rays and/or optical regimes.
Whether the obscuration of AGN signatures takes place on small scales or is linked to the material in the host galaxies can help us distinguish between galaxy-BH (co-)evolution scenarios.
In these proceedings, we highlight key results concerning the incidence of AGN and of X-ray absorption among star-forming galaxies \cite[(see Juneau \etal\ 2013, hereafter J13, for more details)]{jun13}.

\begin{figure}[h]
\begin{center}
\includegraphics[width=4.9in]{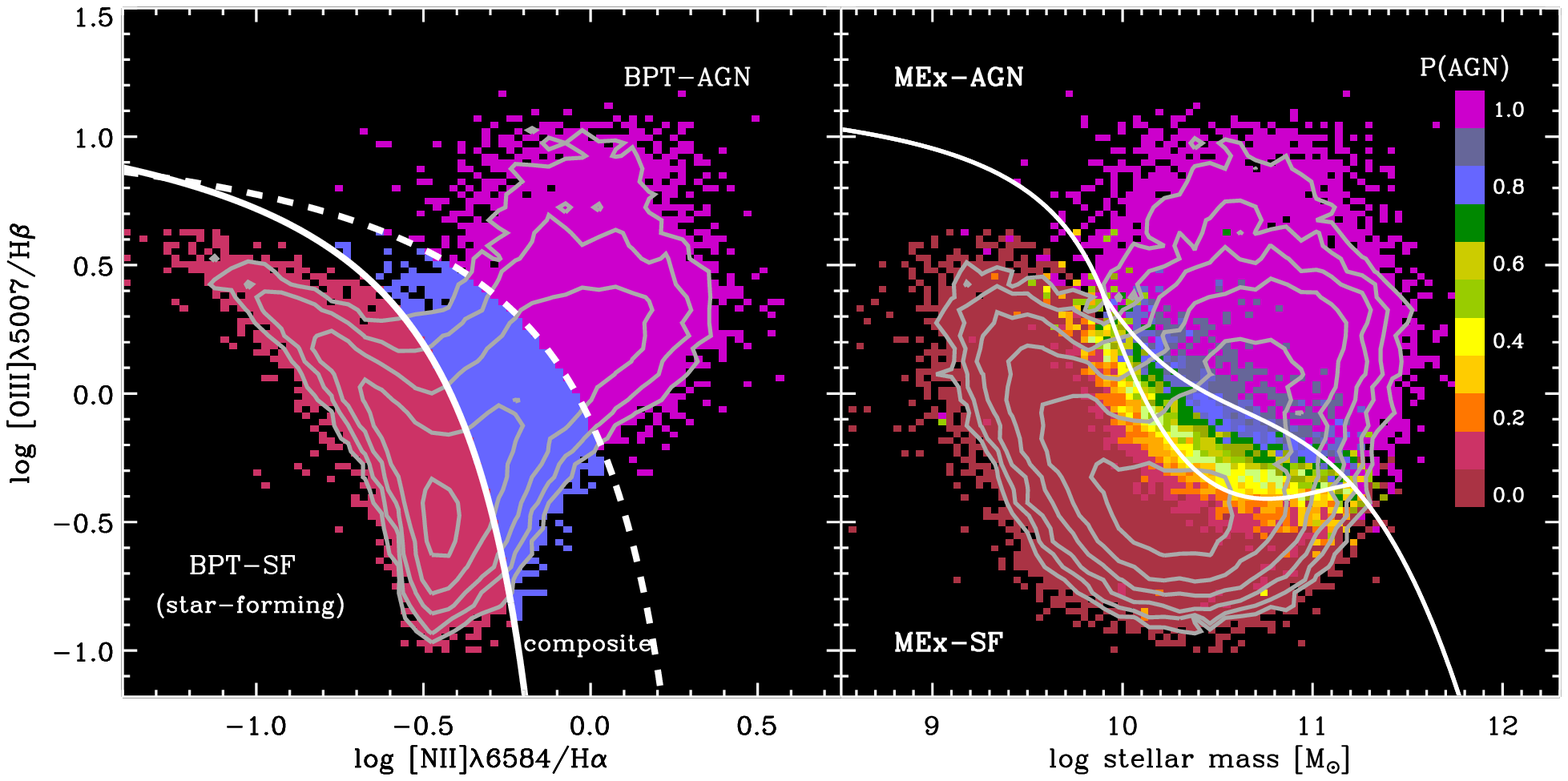}
 \caption{{\it Left.} BPT diagnostic diagram of a calibration galaxy sample at $0.05<z<0.1$ from SDSS.  Star-forming galaxies lie to the bottom left of the \cite{kau03} curve (solid), SF/AGN composites between the curves, and the more extreme AGN above the \cite{kew01} curve (dashed).   
{\it Right.} Mass-Excitation (MEx) diagnostic diagram adapted from \cite{jun11}.  Star-forming galaxies according to the BPT diagram occupy the lower left while AGN lie toward the top right portion of the diagram.  There is a sharp gradient of the fraction of SDSS galaxies hosting AGN across the demarcation lines. See \cite{jun11} for details.
}
   \label{fig1}
\end{center}
\end{figure}

\section{Galaxy Sample and Method}

As described by J13, multiwavelength data are taken from the GOODS and AEGIS surveys. We select star-forming galaxies in the far-IR at 70$\mu$m. Because it comes from dust grains heated by young stars, this emission traces the star formation rate (SFR).  The depth of the FIDEL survey covering both the GOODS-N and EGS fields reaches 3.5mJy (3$\sigma$) in Spitzer/MIPS 70$\mu$m. This flux limit is sufficient to probe down to {\it typical} SFRs at the redshifts of interest ($0.3<z<1$), and includes rarer starbursts, with elevated specific SFR (sSFR$\equiv$SFR/M$_{\star}$; see, e.g., \cite[Elbaz \etal\ (2011)]{elb11} for the evolution of sSFR).  

AGN are identified from one of the following four wavebands: (1) X-ray emission [$L_X\equiv L_{2-10keV}>10^{42}$\,erg\,s$^{-1}$ or hardness ratio $HR>-0.1$]; (2) Mid-IR colors using combination of IRAC bands \cite[(Stern \etal\ 2005; J13)]{ste05}; (3) Optical line MEx diagnostic diagram (Fig.~1); (4) Radio excess \cite[(Del Moro \etal\ 2013)]{del13}.  These methods are used to define three physically-motivated categories (see J13 for more details).

{\underline{\it Unabsorbed AGN}}: X-ray selected AGN that do not show signatures of extreme obscuration such as a very low X/\oiii\ ratio.  They may comprise some moderately absorbed cases with $N_H\sim10^{22-23}\,$cm$^{-2}$ as we do not measure $N_H$ directly due to insufficient X-ray number counts to perform X-ray spectral fitting.

{\underline{\it Absorbed AGN}}: X-ray absorption is inferred from the ratio of X/\oiii\ or the identification with an IR method.  IR methods are mostly sensitive to objects with {\it intrinsic} $L_X>10^{43}\,$erg\,s$^{-1}$.  Therefore IR identification while $L_{X}<10^{42}$\,erg\,s$^{-1}$ implies a factor $>10$ dimming of X-ray emission. Otherwise, the intrinsic AGN luminosity is estimated from \oiii$\lambda$5007, and only kept if it corresponds to intrinsic $L_X>10^{42}$\,erg\,s$^{-1}$.  Some AGN in this category are undetected even with 2\,Ms Chandra data (CDF-N).

{\underline{\it Weak AGN}}: AGN only identified with optical lines (MEx), and with weak \oiii. We estimate that the intrinsic $L_X<10^{42}$\,erg\,s$^{-1}$; these BHs could be accreting inefficiently.

In this work, AGN are counted by adding individual probabilities from the MEx method (Fig.~1), 
noting that X-AGN, IR-AGN and radio-AGN are assigned $P_{AGN}=1$.

\begin{figure}
\begin{center}
\includegraphics[width=5.2in]{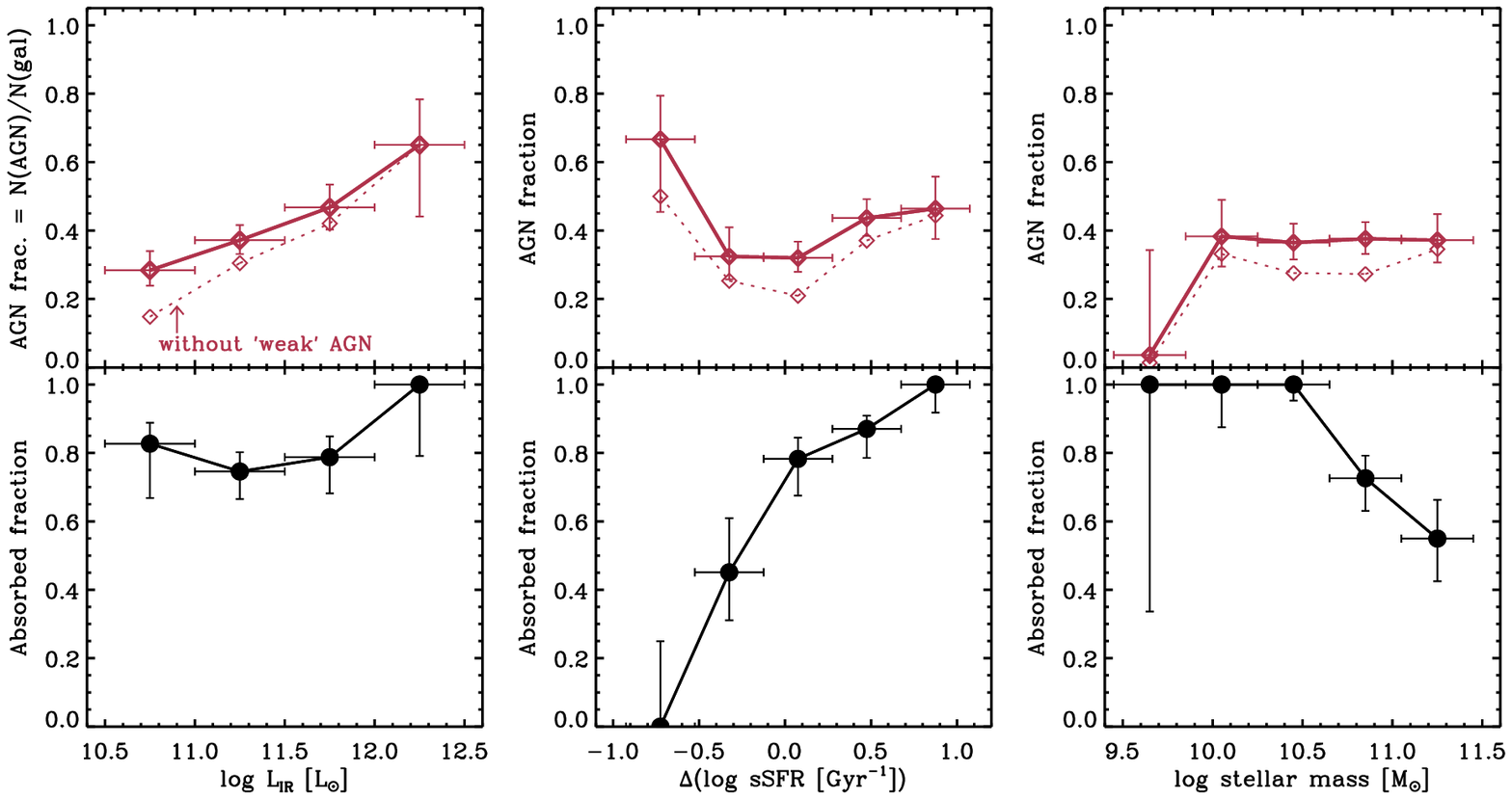}
\caption{{\it Top Row.} Fraction of galaxies with AGN as a function of three galaxy parameters: $L_{IR}$, sSFR, $M_{\star}$.
   We show both the total fraction of AGN (thick symbols), and the fraction without the intrinsically weak AGN (thin symbols). 
   There is a strong increase with $L_{IR}$ but no obvious trends with the other two parameters.
   {\it Bottom Row.} Fraction of AGN that are X-ray absorbed, after removing intrinsically weak systems and non-AGN, 
   as a function of host galaxy parameters: $L_{IR}$, sSFR, $M_{\star}$. There is no trend with $L_{IR}$, but a sharp increase 
   with sSFR.
}
  \label{fig2}
\end{center}
\end{figure}

\section{Results}

{\underline{\it AGN Triggering}}: 
We calculate the fraction of AGN among galaxies as a function of host properties (Top row of Fig.~2).   
We find that the AGN fraction increases with $L_{IR}$, which is proportional to the SFR of the hosts.  This suggests a coeval link between the triggering mechanisms of SMBH growth and star formation, possibly related to the gas reservoirs given that SFRs are proportional to the total mass of dense gas.  On the other hand, we find no obvious trends with the sSFR of the hosts nor with stellar mass. The last result is in contrast with AGN studies based solely on X-ray identification, which tend to be biased toward massive hosts. More trends are investigated in J13 (their Fig.~9).  The overall AGN fraction of 30(37)\% without (with) weak AGN is about twice higher than previous studies based on similarly-selected FIR galaxies in the same intermediate redshift range (see J13 and references therein), thanks to a more complete selection.

{\underline{\it AGN Absorption}}: 
The absorbed fraction among AGN is calculated as: $N_{abs}/(N_{unabs}+N_{abs}$), where $N_{abs}$($N_{unabs}$) is the sum of the AGN probabilities of absorbed (unabsorbed) AGN.  We ignore weak AGN, which may not be absorbed but simply intrinsically faint.  Recalling that the definition of {\it unabsorbed} AGN from the previous section may include moderately absorbed systems, our calculated absorbed fractions may be slightly underestimated.  But because we already find a large absorbed fraction of AGN among star-forming hosts, a larger absorbed fraction would strengthen this result.   The absorbed fraction among AGN is shown as a function of host galaxy properties in Fig.~2.  The relation of AGN absorption with host galaxies are strikingly different from those of AGN triggering.  While there is no clear trend with $L_{IR}$, we find a very sharp increase in the fraction of AGN that are X-ray absorbed with increasing sSFR.  Conversely, the absorbed fraction decreases with increasing stellar mass (on the massive end).  This last trend can result from an evolutionary sequence and/or selection effects, as discussed by J13 (Section~6.3) and summarized below.

\section{Implications}

{\underline{\it AGN Census}}: 
Any single AGN finding method suffers from incompleteness: X-rays can be aborbed by large column densities of gas and suffer from ambiguous identification at low luminosities ($L_X<10^{42}$\,erg\,s$^{-1}$), mid-IR color methods can only pick up dusty AGN and have limited sensitivity usually to $L_X>10^{43}$\,erg\,s$^{-1}$, radio AGN only make up a small fraction of the total AGN population, and optical-line diagnostics like the BPT or MEx diagrams reach to faint accretion rates ($L_X<10^{41}$\,erg\,s$^{-1}$) but can be affected by dust obscuration from the hosts or be blended with SF signatures when weak or moderate AGN are concurrent with an elevated SFR.  Luckily, some methods compensate where others fail, and their combination results in a much more complete sample of AGN.

{\underline{\it AGN Triggering and Duty Cycle}}: 
If galaxies go through active and inactive phases, the fraction of AGN among a population of otherwise comparable galaxies can be interpreted as the duty cycle.  We find that it depends most strongly on the $L_{IR}$ ($\propto$ SFR) of their hosts.  This result implies that the total gas mass dictates the likelihood that a galaxy fuels its central SHBM, a conclusion previously reached for low-redshift galaxies, and recently supported at higher redshift by, e.g., \cite{ros13}.

{\underline{\it Modes of AGN Obscuration}}: 
We found a striking trend between the likelihood of strong X-ray absorption and galaxy sSFR. In contrast, no trend was found with the SFR.  This implies that rather than the total gas mass, the important physical quantity is the gas fraction and/or the gas geometry, both traced to some extent by the sSFR \cite[(Sargent \etal\ 2013)]{sar13}. 
Thus, gas-rich galaxies, and especially those with compact gas reservoirs, are more likely to absorb X-rays from a central AGN. This could be due to either the interstellar gas of the host, or to a multi-scale connection between gas at large scales and the torus at very small scales \cite[(Hopkins \& Quataert 2010)]{hop10}.
Our findings suggest that we need to revise the AGN unification model, in agreement with several contributions made to this IAU symposium (e.g., Ueda, Levenson, Treister, Ridgway, Alonso-Herrero).

{\underline{\it Possible connection with major mergers}}: 
The incidence of AGN is related to the SFR (total gas mass) of their hosts, while the AGN absorption is related to the sSFR (gas fraction or gas geometry).  Major galaxy mergers could explain the most extreme cases because the majority of starbursts in massive galaxies ($>0.3-0.5$\,dex above the average sSFR given their mass and redshift), are experiencing such an event \cite[(Kartaltepe \etal\ 2012)]{kar12}. Another candidate mechanism are gas-rich unstable disks, which are likely to both fuel and obscure their central SMBH \cite[(Bournaud \etal\ 2011, 2012)]{bou11,bou12}.  Large-scale disk instabilities would explain moderate systems, as they do not induce as extreme sSFRs. 


\end{document}